\documentclass{article}
\usepackage{spconf,amsmath,graphicx, amsfonts}
\usepackage[table,xcdraw]{xcolor}
\usepackage[belowskip=-8pt,aboveskip=2pt]{caption}
\title{How Can We Make GAN Perform Better in Single Medical Image Super-Resolution? A Lesion Focused Multi-Scale Approach}
\name{Jin Zhu$^{\star }$ \qquad Guang Yang$^{\dagger \ddagger}$ \qquad Pietro Lio$^{\star}$ }
\address{$^{\star}$ The Computer Laboratory, University of Cambridge, Cambridge, CB3 0FD, UK \\
    $^{\dagger}$ Cardiovascular Research Centre, Royal Brompton Hospital, London, SW3 6NP, UK \\ 
    $^{\ddagger}$ National Heart and Lung Institute, Imperial College London, London, SW7 2AZ, UK }

\begin{document}
\ninept
\maketitle

\begin{abstract}


\noindent Single image super-resolution (SISR) is of great importance as a low-level computer vision task. The fast development of Generative Adversarial Network (GAN) based deep learning architectures realises an efficient and effective SISR to boost the spatial resolution of natural images captured by digital cameras. However, the SISR for medical images is still a very challenging problem. This is due to (1) compared to natural images, in general, medical images have lower signal to noise ratios, (2) GAN based models pre-trained on natural images may synthesise unrealistic patterns in medical images which could affect the clinical interpretation and diagnosis, and (3) the vanilla GAN architecture may suffer from unstable training and collapse mode that can also affect the SISR results. In this paper, we propose a novel lesion focused SR (LFSR) method, which incorporates GAN to achieve perceptually realistic SISR results for brain tumour MRI images. More importantly, we test and make comparison using recently developed GAN variations, e.g., Wasserstein GAN (WGAN) and WGAN with Gradient Penalty (WGAN-GP), and propose a novel multi-scale GAN (MS-GAN), to achieve a more stabilised and efficient training and improved perceptual quality of the super-resolved results. Based on both quantitative evaluations and our designed mean opinion score, the proposed LFSR coupled with MS-GAN has performed better in terms of both perceptual quality and efficiency.



\end{abstract}


\begin{figure*}[t]
    \centering
    \includegraphics[width=\textwidth]{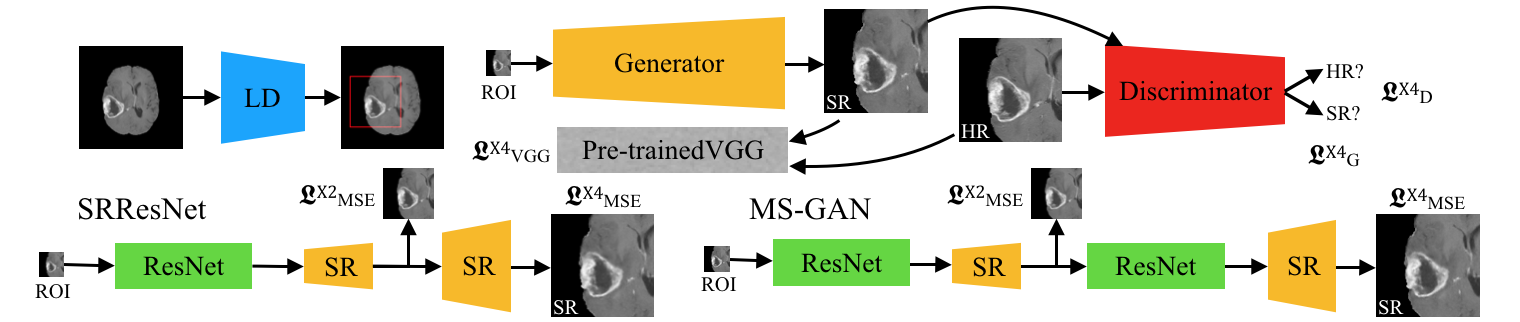}
    \caption{Our proposed lesion focused multi-scale super-resolution neural networks. $\mathrm{LD}$ is a pre-trained lesion detection neural network. Two architectures of generators (SRResNet and MS-GAN) are tested. Adversarial loss, MSE loss and perceptual loss are also displayed.}
    \label{fig:lfmsgan}
\end{figure*}

\section{Introduction}
\label{sec:intro}


Images with high resolution (HR) are greatly in demand for many real applications, e.g., medical images for clinical tasks, geographic information systems, security video surveillance and others \cite{trinh2014novel}. However, the resolution and quality of the images are normally limited by the imaging hardware \cite{moses2011fundamental}, effectiveness and costs. Medical images  are strongly desirable with HR, because they provide crucial details of the anatomical, physiological, functional and metabolic information of patients. In addition to the potential restrictions of the imaging hardware aforementioned, medical images are more susceptible by the health limitations (e.g., ionising radiation dose of using X-ray) and acquisition time limitations (e.g., Specific Absorption Rate limits of using MRI). Moreover, movements due to patients' fatigue and organs pulsation further degrade image qualities and result in lower signal-to-noise ratio (SNR) images. Low resolution (LR) medical images with limited field of view and degraded image quality could reduce the visibility of vital pathological details and compromise the diagnostic accuracy and prognosis \cite{yang2016combined,yang2016super}.

Research studies have shown that instead of optimising hardware settings and imaging sequences, image super-resolution (SR) provides an alternative and relatively cheaper solution for spatial resolution enhancement. Compared to conventional interpolation methods, these SR methods tend to provide better SR outputs with higher SNR and less blurry effects due to the information from multiple LR images or LR-HR image pairs. Reconstruction based SR algorithms have been proven their effectiveness by recovering the HR output with fusing multiple LR images \cite{yang2010exploiting}. However, this type of methods is time-consuming, and the required multi-view LR images are not always available in medical image applications \cite{kang2013self}. Learning based SR methods attract more and more attention now due to their better performance. Simply speaking, this type of SR methods learn a mapping function between LR-HR training pairs (whole images and patches), and apply this mapping to a single testing image to achieve the SR results, namely single image super-resolution (SISR) \cite{trinh2014novel,chang2004super,yang2012coupled}. . 

Recently, deep learning based SISR methods \cite{tai2017image, lim2017enhanced, Hui_2018_CVPR, Zhang_2018_CVPR,deng2018enhancing,ledig2017photo} have boosted the performance of the super-resolved HR images mainly owe to the development of the computing power and the available big data. For example, SRGAN \cite{ledig2017photo}, which is developed based on a Generative Adversarial Network (GAN), has demonstrated perceptually better results compared to other deep residual network \cite{He2015} based SISR methods \cite{Zhang_2018_CVPR,tai2017image,lim2017enhanced}. More recently, Deng \cite{deng2018enhancing} proposed an multi-channel method for SISR which enhanced the objective and perceptual qualities separately. Lai et al. \cite{lai2018fast} incorporated a Laplacian pyramid SR network to progressively super-resolve the sub-band residuals of HR images at multiple pyramid levels. Although these GAN based methods works well on natural images, they are limited for medical images. These pre-trained models using natural images may synthesise unrealistic patterns which could affect the clinical interpretation and diagnosis. Moreover, input LR medical images with lower SNR can intrinsically undermine the performance of the GAN based methods. Thus, SISR for medical images is still an open and challenging problem \cite{han2018cascaded,chen2018efficient}.

In this study, a novel lesion focused SISR method (LFSR) is developed to generate perceptually more realistic SR results and also avoid introducing non-existing features into the lesion area. Because the vanilla GAN architecture may suffer from unstable training and collapse mode problems, newly proposed Wasserstein GAN (WGAN) and WGAN with Gradient Penalty (WGAN-GP) are also tested and compared. Based on our findings, we proposed an advanced multi-scale GAN (MS-GAN) with LFSR to achieve a more stabilised and efficient training procedure and improved perceptual quality of super-resolved results. The validation has been done on MRI images acquired for brain tumour patients using both quantitative metrics and a designed mean opinion score (MOS).

\section{Methods}
\label{sec:methods}

\begin{figure*}[t]
    \centering
    \includegraphics[width=\textwidth]{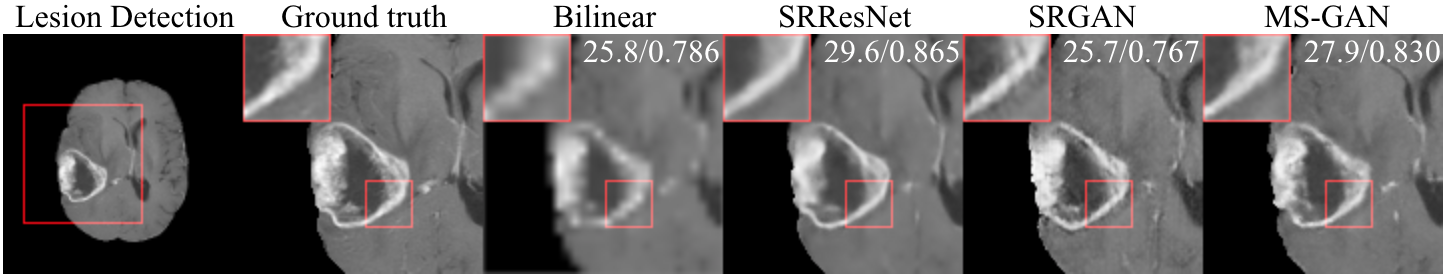}
    \caption{The detected ROI and the predicted SR images, PSNR/SSIM are also displayed.}
    \label{fig:methods}
\end{figure*}
\begin{figure*}[t]
    \centering
    \includegraphics[width=\textwidth]{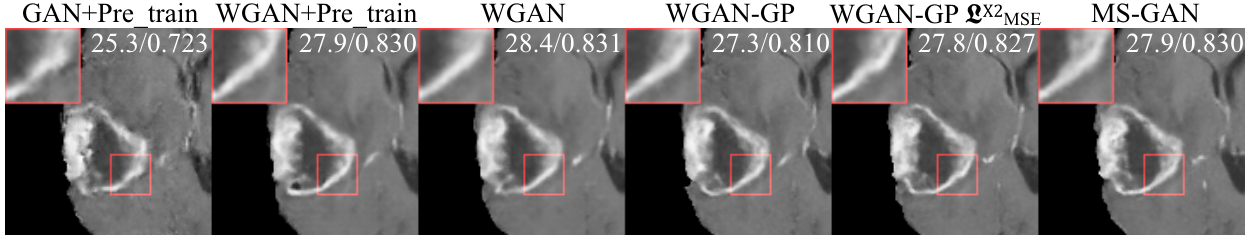}
    \caption{The generated SR images of LFSR with GAN variations and MS-GAN. PSNR/SSIM are also displayed.}
    \label{fig:gans}
\end{figure*}

\subsection{Generative Adversarial Network (GAN)}
\label{ssec:gan}
The originally proposed vanilla GAN \cite{gan} contains a generator $\mathrm{G}$ and a discriminator $\mathrm{D}$ to be trained synchronously via competing with each other. In this work, $\mathrm{G}$ aims to generate as realistic as possible SR images to fool $\mathrm{D}$ and $\mathrm{D}$ aims to distinguish the SR images from real HR images that can be described as:
\begin{equation}
    \hat{\mathrm{G}}, \hat{\mathrm{D}} = \operatorname*{min}_{\mathrm{G}}\operatorname*{max}_{\mathrm{D}}\mathbb{E}_{I_\mathrm{HR}}[log\mathrm{D}(I_\mathrm{HR})] + \mathbb{E}_{I_\mathrm{LR}}[1-log\mathrm{D}(\mathrm{G}(I_\mathrm{LR}))]. 
\end{equation}
where $I_\mathrm{LR}$ and $I_\mathrm{HR}$ are full size LR and HR images, and $\mathbb{E}$ is the expectation of the $D$'s positive outputs (i.e. input is HR ground truth).

However, the vanilla GAN suffers from unstable training, collapsed mode and difficulties in hyper-parameters tuning. Thus, Wasserstein GAN (WGAN) \cite{wgan} has proposed to replace the noncontinuous divergence in $\mathfrak{L}_{\mathrm{D}}$ with the Wasserstein-1 distance:
\begin{equation}
    \hat{\mathrm{G}}, \hat{\mathrm{D}} = \operatorname*{min}_{\mathrm{G}}\operatorname*{max}_{\mathrm{D}\in\mathfrak{D}}\mathbb{E}_{I_\mathrm{HR}}[\mathrm{D}(I_\mathrm{HR})] - \mathbb{E}_{I_\mathrm{LR}}[\mathrm{D}(\mathrm{G}(I_\mathrm{LR}))]  
\end{equation}

In order to enforce the constraint of 1-Lipschitz functions $\mathfrak{D}$, weights clipping is also introduced to ensure all the weights of the $\mathrm{D}$ in a compact range $[-c, c]$. WGAN has shown the advantages to ease the training, increase generative diversity, and promote model flexibility. However, the weights clipping was also found problematic in real applications. Thus, Gradient Penalty \cite{wgangp} was proposed and added into the $\mathfrak{L}_\mathrm{D}$, i.e., WGAN-GP.

\subsection{Single Image Super-Resolution (SISR)}
\label{ssec:lfsr}

The SISR aims to find a transformation to map the distribution of the LR images  $P_\mathrm{LR}$ into the distribution of the HR images  $P_\mathrm{HR}$. A dataset of LR-HR pairs is given as samples of $P_\mathrm{LR}$ and $P_{HR}$ to derive the unknown transformation in a supervised learning manner. In this work, we use deep neural networks to estimate a generated SR distribution $P_\mathrm{SR}$, which has the same dimension as $P_\mathrm{HR}$, and make $P_\mathrm{SR} \approx P_\mathrm{HR}$ according to the chosen metrics. However, in practice, it is challenging to achieve this goal when $P_\mathrm{SR}$ and $P_\mathrm{HR}$ are in a high-dimensional space, because it has very low possibility for them to be overlapped. Thus, we propose two strategies to tackle this problem: lesion focused SISR (LFSR) and multi-scale GAN based SISR (MS-GAN) (Fig.\ref{fig:lfmsgan}). 

\subsubsection{Lesion Focused SISR (LFSR)}
\label{sssec:lfsr}
Firstly, We propose LFSR, which contains a lesion detection neural network $\mathrm{LD}$ and a deep residual super-resolution neural network $\mathrm{SRResNet}$ \cite{ledig2017photo}. $\mathrm{LD}$ aims to detect a region of interest (ROI) of the lesions or abnormalities, e.g., brain tumours in our current study, denoted as $I_\mathrm{lr}$ and $I_\mathrm{hr}$ from the full size LR and HR images $I_\mathrm{LR}$ and $I_\mathrm{HR}$ before we applying the $\mathrm{SRResNet}$, i.e., $I_\mathrm{lr, hr}  = \mathrm{LD}(I_\mathrm{LR, HR})$.

By using $\mathrm{LD}$, the LR-HR image distributions $P_\mathrm{LR} $ and $P_\mathrm{HR}$ are down-scaled as $P_\mathrm{lr}$ and $P_\mathrm{hr}$ in a lower dimensional space. Since only these ROIs are interested in clinical studies, the dimension reduction retains most of the meaningful information from the original full size images. This benefits the training of generative SR models in three aspects: (1) it significantly reduces the training cost of the GAN for the SR task, with a huge reduction of parameters need to be trained; (2) it results in SR images with better perceptual qualities via replacing the estimation of transformation from $P_\mathrm{LR}$ to $P_\mathrm{SR}$ with a much simpler one from $P_\mathrm{lr}$ to $P_\mathrm{sr}$; and (3) the excluded regions will not enter the training process and less artefacts will be synthesised.

\subsubsection{Multi-Scale GAN Based SR (MS-GAN)}
\label{sssec:mssr}
The original SRResNet generates SR images by solving:
\begin{equation}
    \hat{\mathrm{G}}= \operatorname*{min}_{\mathrm{G}} \mathfrak{L}_{\mathrm{SR}}(\mathrm{G}(I_\mathrm{lr}), I_\mathrm{hr})  
\end{equation}
where $\mathfrak{L}_\mathrm{SR}$ can be any predefined loss function. In this paper, we use pixel-wise mean-square-error (MSE) and the VGG \cite{vgg19} based perceptual loss: 
\begin{equation}
    \mathfrak{L}_{MSE} (I_{sr}, I_{hr}) =\frac{1}{W\times H}\sum_{i,j\in W, H}(I_{sr}[i, j]-I_{hr}[i, j])^{2}
\end{equation}
\begin{equation}
 \mathfrak{L}_{VGG} (I_{sr}, I_{hr}) = \mathfrak{L}_{MSE} (V_{l}(I_{sr}), V_{l}(I_{hr}))
\end{equation}
where $W$ and $H$ are the width and height of $I_{sr}$ and $I_{hr}$, and $V_{l}$ is the $l$th layer feature maps of the pre-trained VGG.

Since the original SRResNet generates only one scale SR images, it is hard to stabilise the optimisation process of SR tasks with higher magnifying factors (e.g. X4 magnification). Thus, we propose a MS-GAN architecture to decompose this difficult problem into a series of simpler sub-problems. Our MS-GAN (Fig. \ref{fig:lfmsgan}) can generate multi-scale SR images, and the higher dimensional images are achieved from the lower dimensional ones. For the X4 SR task, both X2 and X4 SR images are sequentially generated. Since the image quality of X4 outputs is based on the performance of X2 ones, the training procedure becomes:
\begin{equation}
    \hat{\mathrm{G}}= \operatorname*{min}_{\mathrm{G}} (\mathfrak{L}^{X2}_{\mathrm{SR}}(\mathrm{G}(I_\mathrm{lr}, X2), I_\mathrm{dr}) + \mathfrak{L}^{X4}_{\mathrm{SR}}(\mathrm{G}(I_\mathrm{lr}, X4), I_\mathrm{hr}))  
\end{equation}
where $I_\mathrm{dr}$ is the X2 down-sampled version of the $I_\mathrm{hr}$. In this work, we choose $\mathfrak{L}^{X2}_{\mathrm{SR}} = \mathfrak{L}_{MSE}$, and $\mathfrak{L}^{X4}_{\mathrm{SR}} = \mathfrak{L}_{MSE} + \mathfrak{L}_{VGG}$, to avoid introducing non-realistic textures in the early stage. In addition with the adversarial loss of GAN $\mathfrak{L}^{X4}_{G}=- \mathbb{E}[\mathrm{D}(\mathrm{G}(I_\mathrm{LR}))]$, the overall loss function of our generator can be denoted as:
\begin{equation}
    \mathfrak{L}_{\mathrm{SR}} =  \mathfrak{L}^{X2}_{MSE} + \mathfrak{L}^{X4}_{MSE} + \mathfrak{L}^{X4}_{VGG} + \mathfrak{L}^{X4}_{G}  
\end{equation}

\subsection{Data Pre-processing and Experiment Settings}
\label{ssec:pp}
The experiments have been done using the open access BraTS 2018 datasets, which contains MRI images acquired from brain tumour patients. In total, 163 patient datasets were included in our study and they were randomly divided into training (9559 slices) and independent validation (2368 slices) groups. All the images were normalised to zero-mean-unit-variance and the LR images were simulated by downsampling the HR ground truth images. 

All the implementation was using Python 3.5, with TensorFlow \cite{tensorflow2015-whitepaper} and TensorLayer, which is now widely used in solving various medical image analysis problems \cite{dong2017automatic,yu2017deep,yang2018dagan}. All the experiments were performed on a Linux workstation with one NVIDIA TITAN X Pascal GPU and Intel(R) Xeon(R) CPU E5-2630 v4 2.20GHz CPUs. All neural networks were trained and tested on the GPU, and the CPUs were only used for data loading and saving.

For a comparison study, we implemented and tested 6 GAN based variations. Firstly, the $\mathrm{SRResNet}$ \cite{ledig2017photo} based LFSR with the vanilla GAN \cite{gan} was tested with a pre-training of the $\mathrm{SRResNet}$ to stabilise the following training of GAN (i.e., 1. GAN+Pre\_train). Then, we tested the same LFSR coupled with WGAN \cite{wgan}, with and without the pre-training of the $\mathrm{SRResNet}$ (i.e., 2. WGAN+Pre\_train and 3. WGAN). Furthermore, the same LFSR with WGAN-GP \cite{wgangp} was trained with and without $\mathfrak{L}^{X2}_{MSE}$ as an extra term of loss function (4. WGAN-GP and 5. WGAN-GP $\mathfrak{L}^{X2}_{MSE}$). Finally, we tested our proposed LFSR coupled with MS-GAN method (6. MS-GAN). All experiments used the same initial learning rate of $10^{-4}$, which decayed to $10^{-5}$ at the midpoint of the training. Although WGAN and WGAN-GP based methods might converge faster than others, all tested methods were trained for 300 epochs to establish a fair comparison. In addition, we also tested the bilinear interpolation, SRResNet and SRGAN\cite{ledig2017photo} for a comprehensive study.

\subsection{Evaluation Metrics}
\label{ssec:eva}
Conventional Peak SNR (PSNR) and Structural SIMilarity (SSIM) index were used to measure the pixel-wise various and image-wise similarity between generated SR results and ground truth HR images. We also designed and performed a mean opinion score (MOS) based evaluation to quantify the perceptual reality of generated SR images. In this study, 100 validation slices were randomly selected for MOS evaluation. For each slice, there were 1 HR ground truth and 6 SR results corresponding to the 6 GAN based variations we tested. Then, we randomly shuffled these 700 images (including 100 HR ground truths). An MR physicist ($>$6 years experience on brain tumour MRI images) performed blinded scoring for these shuffled images based on a Likert-type scale---0 (non-diagnostic), 1 (poor), 2 (fair), 3 (good), and 4 (great)---depending on the image qualities \cite{yang2018fully,seitzer2018adversarial}: over-smooth (S); motion and other kind of artefacts (A); unrealistic textures (U); and too noisy or low SNR (N). The MOS was then derived by calculating the mean and standard deviation for each method.   


\section{Results and Discussions}
\label{sec:results}


\begin{table*}[t]
\centering
\caption{Quantification comparison using PSNR, SSIM and MOS (the best \it{performance}\rm~ in bold)}
\begin{tabular}{l|lllll|llll|ll}
                 & MOS    & Poor & Fair & Good & Great & S  & A  & U  & N  & PSNR  & SSIM     \\
\hline
GAN+Pre\_train   & 1.80$\pm$0.529 & 25   & 71   & 3    & 1     & 2  & 5  & 92 & 91 & 25.2$\pm$1.88 & 0.715$\pm$0.0670  \\
WGAN+Pre\_train & 3.15$\pm$0.669 & 1    & 13   & 56   & 30    & 19 & 0  & 4  & 3  & 27.0$\pm$1.94 & 0.804$\pm$0.0498 \\
WGAN            & 3.09$\pm$0.694 & 1    & 17   & 54   & 28    & 18 & 3  & 7  & 3  & \textbf{27.2$\pm$1.96} & \textbf{0.806$\pm$0.0499} \\
WGAN-GP         & 3.10$\pm$0.686 & 2    & 13   & 58   & 27    & 3  & 0  & 14 & 14 & 26.7$\pm$1.96 & 0.788$\pm$0.0558  \\
WGAN-GP+$\mathfrak{L}^{X2}_{MSE}$      & 3.15$\pm$0.698 & 1    & 15   & 52   & 32    & 3  & 0  & 17 & 17 & 26.2$\pm$2.00 & 0.786$\pm$0.0577 \\
MS-GAN           & \textbf{3.26$\pm$0.626} & 1    & 7    & 57   & 35    & 5  & 1  & 11 & 8  & 26.7$\pm$1.97 & 0.789$\pm$0.0554 \\ 
\rowcolor[HTML]{C0C0C0} 
Ground Truth     & 3.28$\pm$0.838 & 3    & 16   & 31   & 50    & 1  & 20 & 14 & 9  &            &              
\end{tabular}
\label{tab:rst}
\end{table*}

\begin{figure*}[t]
    \includegraphics[width=\textwidth]{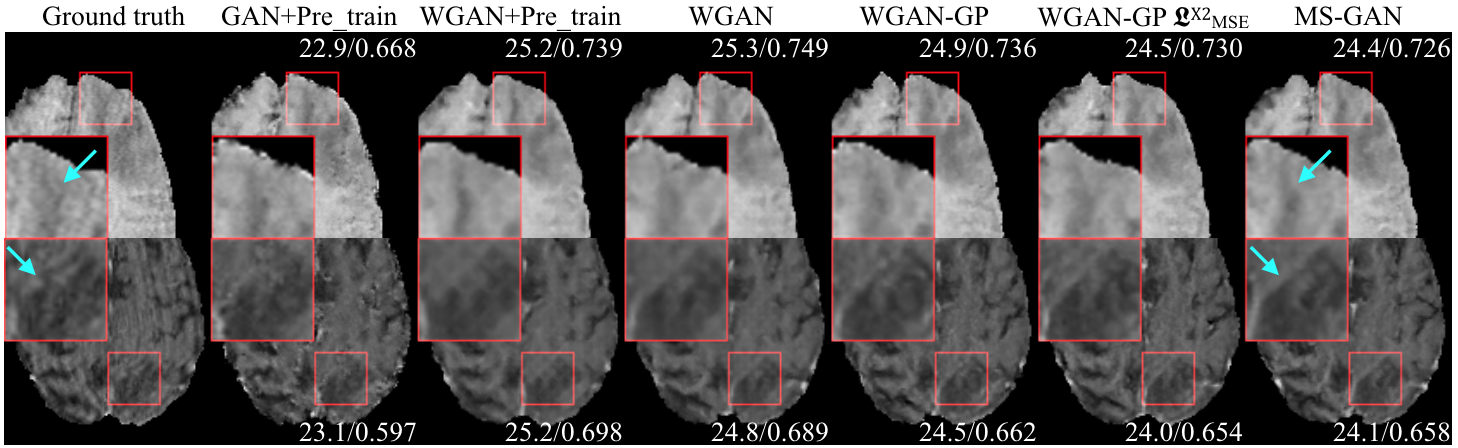}
    \caption{GAN based methods can remove the artefacts in poor image quality ground truth images. Furthermore, our method MS-GAN can enhance the edges and textures of tumour regions. PSNR and SSIM of each result are also displayed.}
    \label{fig:dis}
\end{figure*}

We have tested our proposed LFSR with 6 GAN variations (including the proposed MS-GAN) and different SR image generators for the X4 SR task. It is of note that in order to demonstrate the effectiveness of our MS-GAN, we showed the results of a more challenging X4 SR task, but our proposed methods can also work well with lower magnifying factors (results not shown). Table \ref{tab:rst} tabulates the quantitative results of using LFSR coupled with different GAN models. Except the vanilla GAN produced relatively poor PSNR/SSIM, other GAN variations resulted in similar high PSNR/SSIM. Our MS-GAN method obtained the highest MOS. Figs.\ref{fig:methods} and \ref{fig:gans} show the qualitative visualisation of an example slice. Our MS-GAN achieved high PSNR/SSIM with lesion edge and textural information preserved well. Clearly, compared to the ground truth, vanilla GAN produced noisier SR results. All WGAN based models achieved similar results, but slightly smoother than the results produced by our MS-GAN. Compared to our MS-GAN, although SRResNet yielded higher PSNR/SSIM, perceptually the results were more blurry. SRGAN achieved lower PSNR/SSIM mainly due to the synthesised stripy artefacts in the SR results with less SNR. It is of note that both SRResNet and SRGAN were applied on the whole slice but only the ROIs were evaluated (Fig. \ref{fig:methods}). All the learning based SR methods showed significant improvement over the bilinear interpolation.  

We also evaluated the training and inference efficiency of all methods. The generators influenced on both training and inference costs, while the GAN variations only affected the training cost. Our LFSR with SRResNet \cite{ledig2017photo} and the vanilla GAN \cite{gan} costs 229.6s/epoch for training and 4.04s to generate SR images for the whole validation dataset (2368 slices). According to additional calculation of weight clipping and gradients in WGAN \cite{wgan} and WGAN-GP \cite{wgangp}, the training time increased to 233.8s/epoch and 305.7s/epoch. Moreover, $\mathfrak{L}^{X2}_{MSE}$ also slowed down the training process slightly (314.3s/epoch using WGAN-GP). Finally, because our multi-scale SR generator has more layers, it increases both the training (422.2s/epoch) and inference costs (7.75s for the whole validation dataset). Although the SRResNet took the least cost for each training epoch, it converged much slower than all the others.

Based on our comparative study, there are several interesting findings of the GAN based models: (1) because WGAN and WGAN-GP can stabilise the training better than the vanilla GAN, the pre-training of the generator is no longer necessary; (2) both WGAN and WGAN-GP can provide perceptually more realistic SR than the vanilla GAN, and result in better PSNR/SSIM and significant improvement of the MOS; (3) our proposed LFSR coupled with MS-GAN achieved the most realistic SR with the highest MOS close to the MOS of the ground truth images.

Similar to \cite{ledig2017photo}, our study has also demonstrated the limitations of using PSNR/SSIM as evaluation metrics for medical image SR tasks. Although blurry images are not perceptually realistic enough, they can still achieve relatively high PSNR/SSIM. Comparing all the methods, $\mathrm{SRResNet}$ has achieved the highest PSNR/SSIM, but it has also smoothed out the edge and textural information of the lesion, which are useful and crucial for clinical diagnosis. 

Interestingly, our proposed LFSR with MS-GAN method shows image quality improvement and signal restoration along with the SR. In Fig.\ref{fig:dis}, we can observe that for these two example slices, the ground truth images are with lower SNR and obvious aliasing artefacts (thus, relatively lower MOS). Our MS-GAN method can improve the image quality by boosting the SNR and reducing the artefacts that has resulted in better lesion characteristics (cyan arrows in Fig. \ref{fig:dis}). We can envisage the benefits of our proposed MS-GAN based SISR method for the following clinical image analysis, segmentation and biomarker extraction and characterisation tasks.

\section{Conclusion}
\label{sec:conclusion}

In this study, we propose a novel SISR method to achieve spatial resolution enhancement for the brain tumour MRI images without introducing unrealistic textures. The merits of our work are three-fold: (1) a LFSR has been developed to constrain the deep network to focus on the lesion ROIs, which does not only imitate the clinicians' scrutinization procedure, e.g., enlarge the ROIs, but also dramatically reduce the possible synthesised artefacts from the organs beyond the lesion areas; (2) a comparison study has been carried out to test vanilla GAN with newly proposed WGAN and WGAN-GP to seek possible better GAN based solutions for a more stabilised and efficient training that can yield an improved perceptual quality for the super-resolved results; (3) based on the promises of LFSR and more advanced GAN architectures, a novel MS-GAN model has been developed to tackle the challenges of SISR for medical images especially for the more tricky cases with X4 magnification. In addition to the widely used quantitative metrics (PSNR/SSIM), we also propose the MOS that incorporates experts' domain knowledge for the evaluation of the medical image SR results. Results have shown that our proposed LFSR with MS-GAN can achieve efficient SISR for brain tumour MRI images and we can envisage such models to be successfully applied for a wider range of clinical applications.     

\section{Acknowledgements}       
 
Jin Zhu's PhD research is funded by China Scholarship Council (grant No.201708060173). Guang Yang is funded by the British Heart Foundation Project Grant (Project Number: PG/16/78/32402).




\bibliographystyle{IEEEbibNew}
\bibliography{IEEEabrv,refs}

\end{document}